\documentstyle[11pt]{article}
\newlength{\dinwidth}
\newlength{\dinmargin}
\newlength{\extraspace}
\newlength{\extraspaces}
\newcommand{\be}{\begin{equation}
\addtolength{\abovedisplayskip}{\extraspaces}
\addtolength{\belowdisplayskip}{\extraspaces}
\addtolength{\abovedisplayshortskip}{\extraspace}
\addtolength{\belowdisplayshortskip}{\extraspace}}
\newcommand{\ee}{\end{equation}}
\newcommand{\bdm}{\begin{displaymath}
\addtolength{\abovedisplayskip}{\extraspaces}
\addtolength{\belowdisplayskip}{\extraspaces}
\addtolength{\abovedisplayshortskip}{\extraspace}
\addtolength{\belowdisplayshortskip}{\extraspace}}
\newcommand{\edm}{\end{displaymath}}
\renewcommand{\thefootnote}{\fnsymbol{footnote}}
\def\simlt{\mathrel{\lower2.5pt\vbox{\lineskip=0pt\baselineskip=0pt
           \hbox{$<$}\hbox{$\sim$}}}}
\newcommand{\beq}{\begin{equation}}
\newcommand{\eeq}{\end{equation}}
\newcommand{\bea}{\begin{eqnarray}}
\newcommand{\eea}{\end{eqnarray}}
\newcommand{\ts}{\thinspace}
\newcommand{\semi}{; \\}
\newcommand{\pr}{Phys.\ Rev.\ }

\newcommand{\prl}{Phys.\ Rev.\ Lett.\ }
\newcommand{\np}{Nucl.\ Phys.\ {\bf B}}
\newcommand{\pl}{Phys.\ Lett.\ {\bf B}}
\newcommand{\rmp}{Reviews of Modern Physics\ }

\newcommand{\trho}{\rho_T}

\newcommand{\ppbar}{p{\overline p}}
\newcommand{\gev}{\ts \rm GeV }

\newcommand{\glu}{\cal G}
\newcommand{\epem}{e^+e^-}

\newcommand{\wpwm}{W^+W^-}
\newcommand{\zeenot}{Z^0}

\newcommand{\ewsb} {electroweak symmetry breaking\ }
\begin{document}
\begin{titlepage}
\begin{flushright}
UTPT-95-22\\
hep-ph/9511269
\end{flushright}
\vspace{24mm}
\begin{center}
\Large{{\bf Glueballs in Strongly Interacting Theories\\
at the Electroweak Scale}}
\end{center}
\vspace{5mm}
\begin{center}
M.~V.~Ramana\footnote{e-mail address:
ramana@medb.physics.utoronto.ca}\\ {\normalsize\it Department of
Physics}\\ {\normalsize\it University of Toronto}\\ {\normalsize\it
Toronto, Ontario,}\\ {\normalsize CANADA, M5S 1A7}
\end{center}
\vspace{2cm}
\thispagestyle{empty}
\begin{abstract}

Several proposals for dynamical electroweak symmetry breaking and
fermion mass generation involve strong gauge interactions with a
characteristic scale of a few hundred GeV. The detection of the
glueballs which should occur in such models would be a indication of
the non-Abelian nature of the gauge theories operating at this
scale.  We discuss signatures for these particles.

\end{abstract}
\end{titlepage}
\newpage

\renewcommand{\thefootnote}{\arabic{footnote}}
\setcounter{footnote}{0}
\setcounter{page}{2}

Glueballs have been a long-time prediction in QCD, where they are
believed to occur as a result of the gauge force becoming strong
\cite{who}.  There have been several claims that there is
experimental evidence for their presence \cite{expt}. The best
theoretical estimates of the glueball spectrum in QCD come from
lattice studies \cite{lattice}. These indicate that the lightest
glueballs are scalars with the pseudoscalar and tensor ones being
somewhat heavier.  However the situation in QCD is somewhat confused
because of two features. The first is the presence of $\overline Q Q$
mesons in the same mass range and with the same quantum numbers as
glueballs. This leads to mixing between these states. The second is
the presence of light pseudoscalar mesons.  Glueballs have a large
width due to decays into these light mesons, thus making them broad
states which are difficult to identify.

Dynamical theories of electroweak symmetry breaking are an alternative
to theories involving fundamental scalars. These attempt to explain
the breakdown of the electroweak symmetry by postulating a set of
fermions interacting through a new asymptotically free, strong gauge
force. As in QCD, when the gauge coupling becomes large, the chiral
symmetries of the fermions are spontaneously broken and all the
strongly interacting particles become confined so that all free states
are gauge singlets. Thus in these theories the analog of glueballs
must surely exist. Detection of these glueballs, perhaps in
conjunction with the mesons of these theories, would be a clear
indication of the strong gauge interactions involved in electroweak
symmetry breaking.

The earliest proposal for a dynamical mechanism which effects \ewsb is
technicolor\cite{tc}. This is based on essentially scaling up QCD to
the electroweak scale. In such theories, it would be difficult if not
impossible to detect the analog of glueballs, the techniglueballs, in
collider experiments.  However these theories have problems with
generating realistic fermion masses while suppressing flavor changing
neutral currents and hence are not very interesting to begin with.

In technicolor theories, in order to utilize the chiral symmetry
breaking of technifermions to also break the chiral symmetries of
quarks and leptons {\it explicitly}, one introduces extended
technicolor interactions which couple fermions to technifermions and
are based on some gauge group $G_{ETC}$ \cite{etc,savas}. Walking
technicolor, which enhances the fermion condensate, has been advocated
as a solution to the problem of large flavor changing neutral
currents\cite{etc} in such theories \cite{holdom,techniodor,wtc}.
Besides allowing larger fermion masses, walking also increases the
masses of pseudogoldstone bosons. However since the dynamics of these
theories is quite unlike QCD, there are no reliable ways of estimating
the masses of the bound states in these theories.

The two problems present in QCD may be substantially reduced in the
case of walking theories.  This is because the increase in the fermion
condensate also increases the masses of $\overline Q Q$ mesons (see
\cite{multi,first}) thus raising the technimeson spectrum relative to
the techniglueball spectrum.  Thus the first problem is somewhat
ameliorated, if not wholly circumvented. The increase in the
pseudogoldstone boson mass closes off or kinematically suppresses
techniglueball decay into pseudogoldstone boson pairs. This results in
narrow techniglueball states which are more amenable to detection.

Further attempts in generating large fermion masses involved the ideas
of strong extended technicolor\cite{tom,stetc} and multiscale
technicolor \cite{multi}. The strong ETC picture was suggested as a
means of achieving larger fermion masses, needed to accomodate the
observed top quark mass, while keeping flavor changing neutral
currents within acceptable bounds.  Because of the contribution to the
$W$ and $Z$ masses from extended technicolor interactions, this also
requires a lower technicolor confinement scale\footnote{ For an
example of an explicit low-scale technicolor model, see \cite{king}.},
thus resulting in lighter techniglueballs, with perhaps masses below
100 GeV.

The idea of multiscale technicolor is that fermions transforming under
different representations of the technicolor gauge group should
condense at different scales \cite{raby,kogut,marciano}, thus giving
rise to several scales naturally \cite{multi}. This was advocated as a
means of achieving a walking technicolor coupling while limiting the
number of ETC representations, which is necessary in order to avoid
light axion-like particles (see \cite{etc}).  The same idea has also
risen in attempts to produce a large splitting between techniquarks
and technileptons so as to explain the fermion mass spectrum
\cite{split} and in attempts to produce a large top-bottom mass
splitting while limiting the technicolor contributions to the $\rho$
parameter \cite{revenge}. Once again the constraint of reproducing the
observed $W$ and $Z$ masses requires that the lightest of the scales
in such a theory be fairly low \cite{first}, and the corresponding
glueballs would have masses of the order of a few 100 GeV or less.

Another example of new strongly interacting physics at such low scales
has been suggested by \cite{meta}. This is a mechanism for generating
fermion masses which involves a heavy fourth family and a new sector
of massless fermions which are confined. This strong interaction,
which is termed metacolor in \cite{meta}, would also result in light
glueballs.

Thus we see that there are several scenarios which involved light
glueballs from new strong interactions. In this work we will be
concerned with the signatures of these objects at particle colliders.
We will focus on glueballs in technicolor theories, but much of this
phenomenology is relevant to other theories such as metacolor.

Before moving onto the phenemenology, it may be appropriate to point
out other works which consider glueballs in strongly interacting
theories other than QCD. The earliest instance that has come to our
notice is the attempt in \cite{odor} to explain certain anomalous
events in UA(1) on the basis of a new strong interaction -- odor.  The
new interaction was not involved with electroweak symmetry breakdown.
With the disappearance of the anomalies in $\zeenot$ decays, there has
been no attempt to revive the idea of odor nor to study its
phenomenology. The next mention of glueballs is in \cite{techniodor}
where it was pointed out that the splitting between the chiral
symmetry breaking scale and confinement scale increases in theories
with a walking coupling constant. The last mention of light
techniglueballs is in \cite{tom} where the impact of ETC interactions
on chiral symmetry breakdown is studied.

If we assume that the techniglueballs are below the two
pseudogoldstone boson pair threshold, then the chief decays will be to
pairs of gauge bosons --- $\wpwm$, $\zeenot\zeenot$, $\glu \glu$ and
$\gamma \gamma$. In some cases, the techniglueballs could be so light
that decays to $W$ and $Z$ bosons are kinematically forbidden. Besides
these, there will also be decays to light quarks. However these are
ETC induced couplings and hence small. We may estimate the widths to
pairs of gauge bosons by extending the sum rules mentioned in
\cite{ellis,novikov,chanowitz} to technicolor. The relevant widths are:
\bea
\Gamma(G_T \to \gamma \gamma) &=& {\alpha^2 R^2 M^3}\over
{144 \pi^3 F_G^2} \\
\Gamma (G_T \to \glu \glu) &=& {\beta_c^2 M^3}\over
{8 \pi^2 \alpha_S F_G^2}\\
\Gamma(G_T \to \wpwm) &=& {M^3\over{16 \pi F_G^2}}
\sqrt{(1-4M_W^2/M^2)}\\
\Gamma(G_T \to ZZ) &=& {M^3\over{32 \pi F_G^2}}
\sqrt{(1-4M_Z^2/M^2)}\eea
where $M$ is the techniglueball mass, $F_G$ the techniglueball decay
constant, $\alpha$ the fine structure constant, $R$ the usual ratio of
cross section for hadron production to muon production in $\epem$
colliders, $\beta_c$ is the usual QCD beta function and $\alpha_S$ is
the QCD coupling constant. Given that even in QCD, one does not have
any measured values for the glueball mass or decay constant, the best
we can do is to assume that they both are of the order of a few
hundred GeV. Under that assumption, for $M \simeq F_G = 200 \gev$, we
get:
\bea
\Gamma(G_T \to \gamma \gamma) &=& 3.6\times10^{-5} \gev \\
\Gamma (G_T \to \glu \glu) &=& 1.7 \times 10^{-1} \gev \\
\Gamma(G_T \to \wpwm) &=& 2.4 \gev \\
\Gamma(G_T \to ZZ) &=& 8.3 \times 10^{-1} \gev\ts. \eea
Thus we see that the techniglueball is relatively narrow. But this
also means producing a techniglueball from gauge boson or gluon or
photon fusion is difficult and hence not a viable production mechanism.

In QCD, one preferred production mechanism for glueballs is the
radiative decay of the $J/\psi$ (see, for example, \cite{jpsi}) and
indeed this is one of the channels which has been studied for all the
recent glueball candidates \cite{expt}. This is because the $J/\psi$
is below the threshold for decay into a pair of charm mesons. Thus
except for the radiative decay $J/\psi \to \eta_c \gamma$, all decays
require the two charm quarks to annihilate. This provides a gluon rich
environment and hence there is a relatively large probabibility for
the decay into glueball states. However due to the presence of light
mesons in QCD such as the pions and the rho mesons, these gluons
hadronize into these states often and thus the $J/\psi$ decays into hadrons
about 86\% of the time \cite{pdb}.

The phenomenology of technirhos in walking technicolor theories
\cite{multi,first} is somewhat similar to that of the $J/\psi$
since they often cannot decay into pairs of technipions. The
difference is that in QCD the presence of light quarks allows the
$J/\psi$ to decay into light mesons as mentioned. These decays are not
allowed in walking technicolor. Hence the important decays of the
$\trho$ are into light quarks (and gluons in the case of the
$\trho^8$) \cite{multi,first}, into leptons and into $\eta_T + \gamma$
or $\eta_T + \glu$(in the case of the $\trho^8$) \cite{topq}. These
are the analogs of $J/\psi$ decays into $\epem$, $\mu^+\mu^-$ and its
radiative decays.  If the $G_T$ is lighter than the $\trho$ then the
analog of the glueball decays of the $J/\psi$ are also allowed and
should be large.  Thus the decays $\trho^8 \to G_T + \glu$ and
$\trho^0 \to G_T + \gamma$ should have significant branching ratios.

That the mass of the $G_T$ be less than that of the $\trho$,
especially the $\trho^8$, is not unreasonable in the case of walking
technicolor theories  because of two reasons. The first is the splitting
between the confinement scale and chiral symmetry breaking scales
is expected to increase in a walking theory  as compared to a theory
like QCD with a running coupling constant \cite{techniodor}. Since the
$\trho$ is a bound state of two techni-fermions, its mass
is related to the dynamical fermion mass, which in turn is related
to the chiral symmetry breaking scale. In QCD this splitting has been
suggested as the reason for the success of the nonrelativistic quark
model \cite{georgi}. The second reason is the
enhancement of ETC contributions to the $\trho$ mass in a walking theory
whereas there are no such contributions to the $G_T$ mass. This
becomes even more likely for strong ETC theories.  In
multiscale technicolor theories, it is quite possible that the
techniglueball is lighter than the technirhos from the heavier scales
even if it is heavier than the lightest scale. It is possible though
that the effects of walking could raise even the masses of the technirhos
of the lightest scale to be above that of the techniglueball.

\begin{table}
\centering
\begin{tabular}{|c|c|c|c|}\hline
\  $m_{\trho}$ \ &\ $m_{G_T}$ \  &\multicolumn{2}{|c|}
{$\sigma(\ppbar \to \trho \to {G_T + \glu})$}\\ [1mm] \hline
225 & 100 &\hspace{0.4cm} 0.32\hspace{0.4cm} & 12.77 \\ [2mm]
300 & 50 & \hspace{0.4cm} 0.16 \hspace{0.4cm} & 7.25 \\[2mm]
350 & 200 & \hspace{0.4cm} 0.03 \hspace{0.4cm} & 1.29 \\[2mm]
\hline
\end{tabular}
\label{mej1}
\caption{Cross sections for the production of $G_T + \glu$ in hadron
collisions. The branching ratio for the decay $\trho \to G_T + \glu$
has been estimated by scaling from the decays of the $J/\psi$.  For
each set of masses, the first column the cross section corresponds to
the cross section at the Tevatron and the second to the cross section
at the LHC. All masses are in GeV and all cross sections are in pb.  }
\end{table}

Since it is known that a $\trho^8$ is copiously produced at hadronic
colliders \cite{ehlq,first}, we can expect that this production mode
would lead to plentiful events with a pair of photons or weak gauge
bosons or gluons reconstructing to the $G_T$ mass and a gluonic jet,
with the invariant mass of the whole event reconstructing to the
$\trho^8$ mass. In Tables~1~and~2, we give some estimates of
production cross sections for the $G_T$ in hadron colliders. To obtain
the numbers in Table~1, we have estimated the $\trho \to G_T + \glu$
branching ratio by scaling up the ratio of widths of the $J/\psi$
decays into $f(1500)+\gamma$ \cite{expt} and its decay into $\epem$.
For the numbers in Table~2, we have assumed the branching ratio to be
$0.1$.  The first set of mass values are inspired by the set~A
parameters of \cite{first} and the second, by some of the values in
\cite{tom}.  However unlike these references, we have chosen $N_{TC} =
3$ --- a larger value would imply a larger production cross section
owing to the increased production of a $\trho^8$ of the same mass.

It is clear that the possibility of observing the $G_T$ through this
production mode is very dependent on the branching ratio of the
$\trho$ to $G_T + \glu$.  However it is encouraging to notice that
even with the much smaller branching ratio obtained by scaling from
$J/\psi$ decay, there are significant rates for the production of the
$G_T$ at the LHC. Observing the techniglueball in such a case, as in
the case of most technicolored resonances, is then a matter of
distinguishing the signal from the background. If the $G_T$ decays
into weak gauge bosons or photons, then the backgrounds are relatively
smaller. However, we note that for lighter techniglueballs, as in the
case of the first two sets in Tables~1~and~2, the $G_T$ cannot decay
into weak gauge bosons and hence must decay into gluons or photons.
The backgrounds to events where the $G_T$ decays into a pair of gluons
are much larger. It may however be possible to distinguish these three
jet events from background QCD processes because two of the jets would
reconstruct to a narrow width $G_T$. Given the large uncertainities in
even estimating the signal rate, we will not discuss the background in
great detail.

\begin{table}
\centering
\begin{tabular}{|c|c|c|c|}\hline
\  $m_{\trho}$ \ &\ $m_{G_T}$ \  &\multicolumn{2}{|c|}
{$\sigma(\ppbar \to \trho \to {G_T + \glu})$}\\ [1mm] \hline
225 & 100 &\hspace{0.4cm} 14.95\hspace{0.4cm} & 587.69 \\ [2mm]
300 & 50 & \hspace{0.4cm} 4.14 \hspace{0.4cm} & 187.95 \\[2mm]
350 & 200 & \hspace{0.4cm} 1.99 \hspace{0.4cm} & 100.40 \\[2mm]
\hline
\end{tabular}
\label{mej2}
\caption{Cross sections for the production of $G_T + \glu$ in hadron
collisions. The branching ratio for the decay $\trho \to G_T + \glu$
has been assumed to be 0.1.  For each set of masses, the first column
the cross section corresponds to the cross section at the Tevatron and
the second to the cross section at the LHC. All masses are in GeV and
all cross sections are in pb.  }
\end{table}

In $\epem$ collisions, one could look for events with a pair of gluons
or weak gauge bosons reconstructing to a $G_T$ and a photon, with the
invariant mass of the whole event reconstructing to the singlet
$\trho^0$ mass. Due to the relatively clean environment of $\epem$
colliders, this signal should be quite distinctive.

Recently an interesting mechanism for the production of QCD glueballs
has been suggested \cite{losalamos}. Their argument is that in $\epem$
collisions at energies somewhat above the threshold, a pair of heavy
quarks would settle into a vector meson state by the emission of a
glueball with a large probability. The analog of this mechanism in the
case of technicolor is the production of a techniglueball in
conjunction with a technirho meson in either $\epem$ or hadron
collisions. This mechanism could be even more effective if there are
excitations of the $\trho$ (i.e. a $\trho^*$) with masses of
approximately $m_{\trho} + m_{G_T}$. The $\trho$ could be either a
color singlet ($\trho^0$) or a color octet ($\trho^8$) in the case of
hadronic collisions.  However one would expect the color octet mode to
be the dominant one.  In $\epem$ collisions, since the initial states
carry no color charges, one cannot produce the octet technirho in
conjunction with a techniglueball.

In hadronic collisions, the production of $\trho + G_T$ followed by
the subsequent decay of the $\trho$ into jets in the case of a color
octets \cite{first} or weak gauge boson pairs in the case of color
singlets \cite{multi} and the decay of the $G_T$ into a $\gamma
\gamma$ pair or $\wpwm$ or $ZZ$ should be a clean signal that can be
easily distinguished from backgrounds due to the resonance structure
of the signal. Likewise in $\epem$ collisions the production of a
color singlet $\trho$ and a $G_T$ followed by the decay of the $\trho$
into weak gauge boson pairs and the decay of the $G_T$ into weak gauge
bosons or two jets ($\glu\glu$) should be a clean signature. In both
cases, requiring the pairs of electroweak gauge bosons or jets or
photons to reconstruct to definite masses should be a useful way of
discriminating signal events from the background. It may be noted that
while the cross section for production of just $\trho$ may be much
larger, the $\trho +G_T$ channel has a more unique signature and thus
it should be much easier to distinguish this from potential
backgrounds.

In conclusion we have suggested a new signature for strongly
interacting theories with characteristic scales of the order of a few
hundred GeV. The detection of glueballs associated with these strongly
interacting theories would be an indication of the non-Abelian nature
of these gauge theories operating at this scale and would be
complementary to the detection of the mesons of these theories which
has been considered by several people. We have provided arguments for
why these should be relatively light and narrow and hence more easily
observable. Two production mechanisms have been suggested; their
relative importance is a function of the detailed dynamics of the
model considered.

\newpage

{\bf Acknowledgements}

I would like to thank U.~Mahanta and T.~Appelquist for suggesting that
I look at signatures for techniglueballs and K.~Lane for urging me to
look for viable production mechanisms and backgrounds. I thank
D.~Kominis for his comments on the manuscript. I would also like to
thank B.~Holdom for a careful reading of the manuscript, his comments,
clarification of several issues and last but not least his
encouragement. This research was supported in part by the Natural
Sciences and Engineering Research Council of Canada.

\end{document}